\documentclass{mem}
\usepackage{natbib}\usepackage{txfonts}\usepackage{balance}
\usepackage{graphicx}
\usepackage[a4paper]{hyperref}
\idline{?}{1}
\begin{document}
\def\teff{$T\rm_{eff }$}
\def\kms{$\mathrm {km s}^{-1}$}

\title{
On the nature of small amplitude peaks \\
in $\delta$ Scuti oscillation spectra
}

\author{
J. \,Daszy\'nska-Daszkiewicz\inst{1,2},
W. A. \,Dziembowski\inst{2,3},
A. A. \, Pamyatnykh\inst{2,4}
          }

  \offprints{J. Daszy\'nska-D.}

\institute{
Instytut Astronomiczny, Uniwersytet Wroc{\l}awski, ul. Kopernika 11,
51-622 Wroc{\l}aw, Poland, \email{daszynska@astro.uni.wroc.pl}
\and
Copernicus Astronomical Center, Bartycka 18,
00-716 Warsaw, Poland
\and
Warsaw University Observatory, Al. Ujazdowskie 4,
00-478 Warsaw, Poland
\and
Institute of Astronomy, Russian Academy of Sciences,
Pyatnitskaya Str. 48, 109017 Moscow, Russia
}

\authorrunning{J. \,Daszy\'nska-Daszkiewicz et al.}

\titlerunning{
Small amplitude peaks in $\delta$ Scuti oscillation spectra
}

\abstract{ The standard assumption in interpretation of stellar
oscillation spectra from photometry is that the excited mode have
low angular degrees, typically  $\ell< 3$. Considering the case of
FG Vir, the $\delta$ Scuti star with the richest known oscillation
spectrum, we show that this assumption is not justified for low
amplitude peaks. The $\ell<3$ identifications have been found for
12 dominant peaks from pulsation amplitudes and phases. However,
we show that for the rest of the peaks (55), whose amplitudes are
typically  below 1 mmag, much higher $\ell$'s are most likely.
 We argue that improving amplitude resolution to the
micromagnitude level, as expected from space observations, is not
likely to be rewarded with a credible mode identifications because
the spectra will be dominated by high-$\ell$ modes of unknown
azimuthal order, $m$.  } \maketitle{}

\section{Introduction}

The argument invoked for limiting possible identification to
$\ell<3$ degrees is based on the fact that, at the same intrinsic
pulsation amplitude, the disc-averaged amplitude is significantly
reduced at higher $\ell$'s. The disc-averaging factor,
$b_{\ell}^{\lambda}$, which involves the limb-darkening law, jumps
between $\ell=2$ and $\ell=3$ (see Fig.\,1). However, the argument
is problematical especially in the case of oscillation spectra
exhibiting peaks differing drastically in amplitudes. There are
many examples of oscillation spectra for opacity driven pulsators
where the amplitudes differ by more than two orders of magnitude.
This difference corresponds to going from $\ell=0$ to $\ell=6$ in
the $b_{\ell}$ value. In addition, if $\ell$ is sufficiently large
so that the geometrical effect dominates then there is the
$(\ell-1)(\ell+2)$ factor in front of $b_{\ell}$ in the expression
for the light amplitudes.

Possibility of detecting high degree modes from photometry was
already discussed by Balona \& Dziembowski (1999), who considered
various types of pulsators on the upper main sequence as well as
stars in the Cepheid instability strip.
\begin{figure}[]
\resizebox{\hsize}{!}{\includegraphics[clip=true]{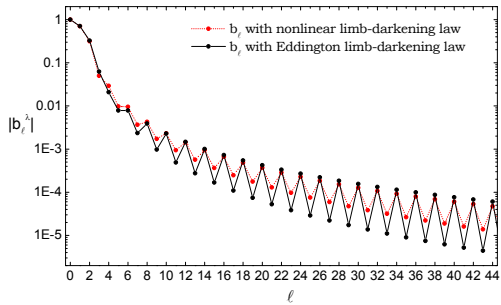}}
\caption{\footnotesize The disc-averaging factor, $b_{\ell}$, as a
function of $\ell$ for two limb-darkening laws. The
nonlinear limb darkening law was calculated at $\log T_{\rm
eff}=3.86$ and $\log g=3.9$.}
 \label{}
\end{figure}

In this short report, we focus on FG Vir star, which is the most
multimodal $\delta$ Scuti pulsator known so far. According to the
recent paper of Breger at al. (2005), there are 67 independent
peaks detected in this star covering the fequency range from about
6 up to about 45 c/d. In order to decipher this rich oscillation
spectrum, we need at least some idea about the spherical harmonic
of the excited mode. We know that high degrees must be involved,
because there is just not enough low degree modes in certain
overdense ranges of the spectrum. Moreover, not all low degree
modes must be excited. Simple nonlinear simulations suggest that
pulsational instability is saturated by a random subset of
unstable modes, which may include only a part of the low degree
modes (Nowakowski 2005).

In the frequency spectrum of FG Vir, twelve modes were detected
both in photometry and spectroscopy (radial velocity). For all of
them the amplitude and phase data lead to $\ell<3$ identification
(Daszy\'nska-Daszkiewicz et al. 2005). Unfortunately, for the
remaining 55 peaks the data do not allow for a unique $\ell$
determination. Below, we present a speculation what $\ell$'s of
those modes are most likely.

\section{Mode instability and saturation}

The range of observed modes in the FG Vir oscillation spectrum
extends from about 6 up to about 45 c/d. Model calculations
predict p-mode instability extending only up to about 30 c/d and
up to $\ell=30$. Above these limits, only f-modes with $\ell$
between 70 and 200 are unstable. The instability ranges for p$_1$
and f-modes are shown in Fig.\,2. For all calculations, we chose
the model with the following parameters: $M=1.85 M_{\odot}$, $\log
T_{\rm eff}=3.8603$, $\log L/L_{\odot}=1.17$ and solar chemical
composition.  The same model was used by Dasz\'nska-Daszkiewicz et
al. (2005) in their analysis of the data on the 12 dominant peaks.
At low degrees, the mode driving arises mainly in the HeII
ionization zone. With increasing the mode degree, the driving
source moves toward  the upper layers and ultimately at $\ell>60$
it takes place mostly in the H ionization zone.
\begin{figure}[]
\resizebox{\hsize}{!}{\includegraphics[clip=true]{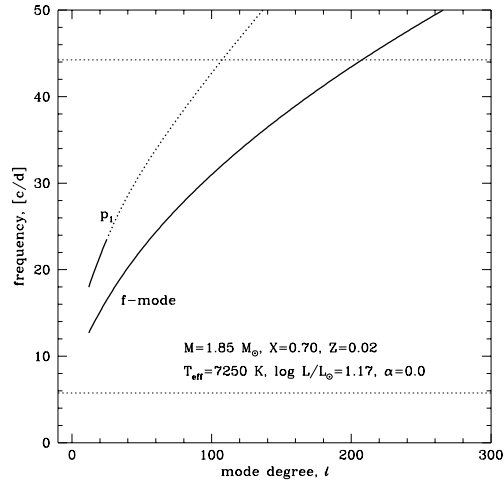}}
\caption{\footnotesize Frequencies of high degree p$_1$ and f modes as
functions of mode degree. The solid line refers to unstable modes,
whereas the dotted line to the stable ones. Two horizontal lines
mark the observed frequency range in FG Vir. } \label{}
\end{figure}

What is the upper limit of mode degree at specified photometric
amplitude? The answer depends on the intrinsic mode amplitude,
$\varepsilon=\!<\delta R/R\!>_{\rm rms}$, and on the amplitude
resolution in the photometric data sets, $A^{\rm crit}$. In Breger
et al. (2005) data it is 0.2 mmag. Unfortunately, we do not have a
credible theory predicting intrinsic amplitudes for multimode
pulsators. Everything what we have at the moment are preliminary
assesments based on simulations made by Nowakowski (2005). His
simulations suggest that the main amplitude limiting effect is not
a resonant mode coupling but rather a collective saturation of the
opacity driving mechanism. If this is a case, then modes in a wide
range of $\ell$ degrees are on equal footing. Except that,  of
course, there are more high $\ell$ modes that low $\ell$ modes.
These simulations also show that the terminal state of pulsation
is a random realization out of many possibilities.

We had to exclude the f-modes with high $\ell$'s from our
consideration because at the 0.2 mmag amplitude, the linear model,
which we used, fails in the H ionization zone. Note, however, that
this problem does not invalidate interpretation of the high
frequency peaks in terms of such modes.

\section{Visibility of modes with high $\ell$}

For our estimates of observable amplitudes, we first assumed that
the intrinsic amplitudes, $\varepsilon$, of all modes are the same
and equal to amplitude of the dominant mode, $\nu_1$, which was
identified from photometric and spectroscopic data as $\ell=1,
m=0$. The value of this intrinsic amplitude, $\varepsilon=0.005$,
was estimated in our recent paper (Daszy\'nska-Daszkiewicz et al.
2005). The adopted inclination angle was $20^{\circ}$, which is
close to the value given by Mantegazza \& Poretti (2002). We
considered modes with $\ell<45$ and $\nu<30$ c/d and all possible
azimuthal orders, $m$.

In Fig.\,3 we plot the number of modes at a given $\ell$ degree
with photometric amplitudes in the $y$ Str\"omgren passband,
exceeding 0.2 mmag. This value corresponds to the amplitude
resolution in FG Vir oscillation spectrum (Breger et al. 2005). We
can see that most modes have harmonic degrees from about $\ell=4$
to $\ell=9$. The rapid decrease of the mode number around
$\ell=27$ results from stabilization of p$_1$ modes. The faster
decline for odd $\ell$ modes is a consequence of much smaller
$b_{\ell}$ factor than for even $\ell$ modes. We presented the
results only for one inclination angle, $i=20^{\circ}$, but the
influence of $i$ is small, except very near to the pole-on
direction ($i=0^{\circ}$).
\begin{figure}[]
\resizebox{\hsize}{!}{\includegraphics[bb= 49 311 484 584,clip=true]{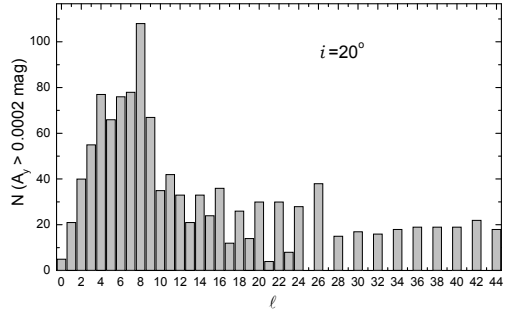}}
\caption{\footnotesize Number of modes with photometric amplitudes
$A_y$ exceeding 0.2 mmag as a function of the spherical harmonic
degree, $\ell$. The values of the adopted intrinsic amplitude and
the inclination angle are $\varepsilon=0.005$ and $i=20$,
respectively. } \label{}
\end{figure}
\begin{figure}[]
\resizebox{\hsize}{!}{\includegraphics[bb=52 377 475
651,clip=true]{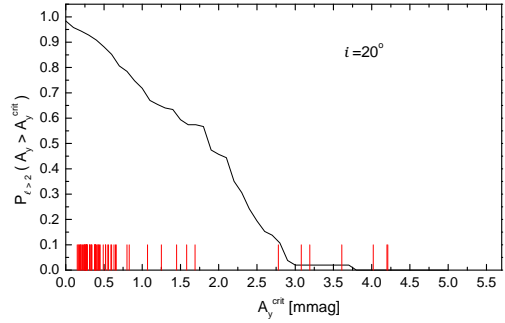}} \caption{\footnotesize Probability
that the modes have $\ell>2$ at different threshold amplitudes.
The assumed values of the intrinsic amplitude and he inclination
angle are $\varepsilon=0.005$ and $i=20^{\circ}$, respectively.
Bars mark the amplitudes of detected modes in FG Vir, except the
dominant one, which has much larger amplitude. } \label{}
\end{figure}
\begin{figure}[]
\resizebox{\hsize}{!}{\includegraphics[bb= 52 289 475
568,clip=true]{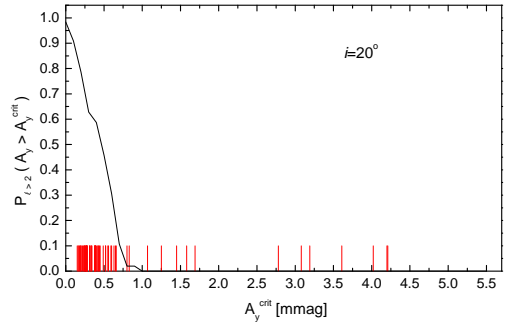}} \caption{\footnotesize The same as in
Fig.\,4 but the intrinsic amplitudes were assumed five times less
($\varepsilon=0.001$). } \label{}
\end{figure}

The number of modes can be directly translated into the
probability. In Fig.\,4 we depict the probability that the
observed modes are $\ell>2$ at different threshold (critical)
amplitudes. In the abscissa, we marked also the photometric
amplitudes of all modes detected in FG Vir except $\nu_1$. We may
see that at below the 2 mmag amplitude, the modes most likely have
$\ell\!>\!2$. The fact that four of such modes we identified as
$\ell\!<\!3$ may indicate either unlikely coincidence or that that
the selection mechanism favours low $\ell$ degree modes. Another
option, which we regard more likely, is that the typical intrinsic
amplitudes are somewhat lower than we have assumed. Fig.\,5 shows
the similar results but obtained upon assuming the intrinsic
amplitude five times less $(\varepsilon=0.001)$. This value
corresponds to the $\nu_2$ peak of FG Vir which we identified as
$\ell=0$. We can see that now the value of
$P_{\ell>2}(A_y>A_y^{\rm crit})$ decreases much more rapidly, but
still most of the peaks have a high probability to be $\ell>2$
modes.

There are many modes in FG Vir oscillation spectrum which are seen
in photometry but not in spectroscopy. What should be the
amplitude of the radial velocity for such modes? For a few of
them, we calculated the radial velocity amplitude at a given
observed amplitude in photometry. The result is presented in
Fig.\,6. We may see that some of such modes should have been
detected by spectroscopy if they had $\ell\!<\!4$. Note also, that
$\ell=3$ mode should be most easily detectable in the radial
velocity. If $\ell$ had been equal 3 all the modes should have
been detected.
\begin{figure}[]
\resizebox{\hsize}{!}{\includegraphics[bb= 36 455 482 728,
clip=true]{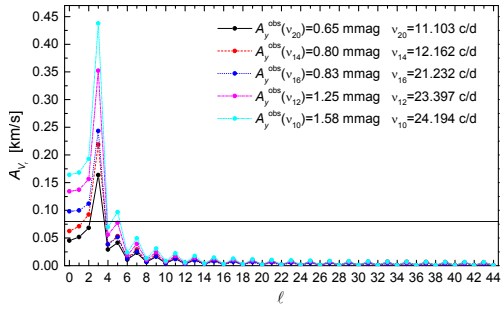}} \caption{\footnotesize The radial
velocity amplitudes implied by the photometric amplitudes, $A_y$,
at various assumed mode degree, $\ell$, for five relatively high
peaks in the FG vir oscillation spectrum. The horizontal line
marks the detection threshold according to Zima (private
communication).} \label{}
\end{figure}
\section{Conclusions}

We have shown that the oscillation spectrum of FG Vir cannot be
explained in terms of low-$\ell$ modes alone. Above frequency of
30 c/d only high degree ($\ell > 70$) f-modes are unstable. Also
below 30 c/d most of the observed modes must have $\ell>2$. Our
simulations, in which we assumed the same intrinsic amplitudes,
showed that the most likely degrees are $\ell>5$.

The immediate conclusion from  our work is pessimistic. $\delta$
Scuti stars seem not a good target for space observations. At the
expected micromagnitude threshold, the oscillation spectra will be
dominated by high degree modes of unknown azimuthal order, mode
identification will be impossible task, thus the frequencies will
not be useful for seismic probing. However, we should not forget
about our underlying assumption that the intrinsic amplitude are
the same for all modes. The assumption may not be justified and we
are curious what will the space data tell us. In any case, these
data will teach us about the mode selection mechanism in multimode
pulsators.

\begin{acknowledgements}
The work was supported by the Polish MNiI grant No. 1 P03D 021 28.
\end{acknowledgements}

\bibliographystyle{aa}

\end{document}